%% file: main.tex
\RequirePackage{fix-cm}
\documentclass[smallextended]{svjour3}       
\smartqed  

\usepackage[normalem]{ulem} 
\usepackage{graphicx}
\usepackage{multirow}
\usepackage{tcolorbox}
\usepackage{hyperref}
\usepackage{caption}
\usepackage{subcaption}
\usepackage{todonotes}

\newcommand{\interviewquote}[2]{\begin{quote}
\footnotesize{\emph{``#1'' }} --- \footnotesize{#2}
\end{quote} }

\begin{document}
\title{Exploring Actions, Interactions and Challenges in Software Modelling Tasks: \\An Empirical Investigation with Students}
\titlerunning{An Empirical Investigation with Students}
\author{Shalini Chakraborty \and Javier Troya \and Lola Burgueño \and Grischa Liebel}
\date{July 2024}

\institute{S. Chakraborty \at
              Reykjavik University\\ Menntavegur 1, 102 Reykjav\'{i}k, Iceland \\ ORCID: 0000-0002-9466-3766 \\
              \email{shalini19@ru.is}
\and 
J. Troya \at
              ITIS Software, Universidad de M\'{a}laga\\ Blvd. Louis Pasteur, M\'{a}laga, 29071, Spain \\ ORCID: 0000-0002-1314-9694 \\
              \email{jtroya@uma.es} 
\and
L. Burgue\~{n}o \at
              ITIS Software, Universidad de M\'{a}laga\\ Blvd. Louis Pasteur, M\'{a}laga, 29071, Spain \\ ORCID: 0000-0002-7779-8810 \\
              \email{lolaburgueno@uma.es} 
\and
G. Liebel \at
              Reykjavik University\\ Menntavegur 1, 102 Reykjav\'{i}k, Iceland \\ ORCID: 0000-0002-3884-815X \\
              \email{grischal@ru.is}      
}

\date{Received: date / Accepted: date}
\maketitle
\begin{abstract}
\textbf{Background:}  Software modelling is a creative yet challenging task. Modellers often find themselves lost in the process, from understanding the modelling problem to solving it with proper modelling strategies and modelling tools. Students learning modelling often get overwhelmed with the notations and tools. To teach students systematic modelling, we must investigate students' practical modelling knowledge and the challenges they face while modelling.  
\textbf{Aim:} We aim to explore students' modelling knowledge and modelling actions. Further, we want to investigate students' challenges while solving a modelling task on specific modelling tools. 
\textbf{Method:} We conducted an empirical study by observing 16 pairs of students from two universities and countries solving modelling tasks for one hour. 
\textbf{Results:} We find distinct patterns of modelling of class and sequence diagrams based on individual modelling styles, the tools' interface and modelling knowledge. We observed how modelling tools influence students' modelling styles and how they can be used to foster students' confidence and creativity. Based on these observations, we developed a set of guidelines aimed at enhancing modelling education and helping students acquire practical modelling skills.
\textbf{Conclusions: } The guidance for modelling in education needs to be structured and systematic. Our findings reveal that different modelling styles exist, which should be properly studied. It is essential to nurture the creative aspect of a modeller, particularly while they are still students. Therefore, selecting the right tool is important, and students should understand how a tool can influence their modelling style. 
\end{abstract}

%
\input{01Introduction}

\input{02RelatedWork}
\input{03Method}

\input{04Results}

\input{05Discussion}
\input{06Conclusion}

\section*{Acknowledgement}
We would like to thank all the students who participated in the study.

\section*{Declarations}
\subsection*{Competing Interests}
The authors declare that they have no conflict of interest.

\subsection*{Data Availability}
We publish the problem descriptions we used for pilot studies 1 and 2 and the main study in the Appendix. We didn't disclose the videos of students modelling for privacy reasons.
\bibliographystyle{plain}
\bibliography{reference}
\appendix
\input{07appendix}
\end{document}

%% file: 01Introduction.tex

\section{Introduction}
Software modelling holds significant promise for enhancing various aspects of software and systems engineering, such as productivity \cite{agner12} and cost efficiency~\cite{kirstan10}. Despite these advantages, the widespread adoption of software modelling across the entire field remains limited \cite{gorschek2014use}. 
Extensive research has explored the reasons behind the limited adoption, revealing issues like subpar code generation, inadequate tool support, and a lack of guidance or training \cite{Forward2010perceptions,Whittle2013Industrials,Whittle2014StateofPractice,Mohagheghi13Help,liebel18sosym,liebel18survey}. Specifically, it has been suggested that engineers are reluctant to embrace modelling because it requires excessive effort and offers little usefulness, a view shaped by their educational background \cite{gorschek2014use,torchiano13}.
Within the educational domain, substantial efforts have been dedicated to understanding how to teach modelling within university curricula, exemplified by works such as \cite{Westphal2019Teaching,Loli2018Teaching,kolovos2016towards,paige2014bad,whittle2011mismatches,Stikkolorum2015LogViz,Liebel2016Impact}. This body of literature often takes the form of proposed course designs \cite{schmidt2014teaching,Westphal2019Teaching}, experiential reports detailing challenges or best practices, particularly concerning tools \cite{Loli2018Teaching,kolovos2016towards,paige2014bad,whittle2011mismatches}, or papers presenting quantitative studies on student opinions \cite{Stikkolorum2015LogViz,Liebel2016Impact}. Collectively, this wealth of knowledge serves as a valuable source of experience and inspiration for the effective teaching of modelling practices.
%
However, research on students' perspectives on modelling needs to provide more detailed explanations, such as specific challenges and benefits regarding modelling assignments, modelling tools, syntax, and semantics. 
Furthermore, experience reports from researchers with a specific focus on software and systems modelling may run the risk of being unrepresentative for instructors lacking such a focus. Given that most students will transition into professional roles post-graduation, it becomes crucial to comprehend their perceptions. It is essential to understand the challenges students currently face and explore the process of modelling that students practice in universities. 

In our prior research \cite{chakraborty2023we}, we engaged in 16 interviews with students and 5 interviews with instructors to explore students' perceived understanding of modelling and the challenges they encountered. The findings of our study indicate that students recognise specific advantages of modelling, particularly in areas where informal models suffice, such as gaining a system overview or conceptual comprehension of the problem domain. However, the majority of the students remains skeptical about the value of modelling in detailed system design. Many students expressed challenges related to issues such as selecting the appropriate notation, determining what aspects to express in the models, and understanding how to apply modelling techniques to unfamiliar domains.
%

In this paper, we dig deeper into students' modelling practices by undertaking an observational study. We conducted two modelling sessions where students participated in pairs and solved modelling tasks. Our goal here is to observe students' modelling actions and interactions during the modelling task and explore the challenges experienced by an individual or pair.

%
The study follows two research questions (RQs):
\begin{itemize}
    \item RQ1: What actions do students take on a modelling tool to solve a task? 
    \item RQ2: What challenges do students face while solving modelling tasks?
\end{itemize}

RQ1 focuses on the actions students undertake within a modelling tool as they solve a given task. Understanding the specific steps students employ in a modelling tool provides valuable insights into their cognitive processes and problem-solving strategies.
RQ2 investigates the challenges students experience while engaging in modelling tasks, including challenges based on students' modelling perspectives, tool knowledge and the communicative dynamics between pairs of students engaged in collaborative problem-solving. 
%


This paper presents findings from an observational study conducted with 32 students from two different universities, Reykjavik University, Iceland, and University of Malaga, Spain.  The students worked in pairs, engaged in solving tasks using two modelling tools, namely MagicDraw\footnote{https://www.magicdraw.com/} and PlantUML\footnote{https://plantuml.com/}.

Our results show that students need more knowledge of structuring and completeness in modelling. While students are confident drawing class diagrams, sequence diagrams still pose a challenge to them. In both cases, students need to improve how they read and process the problem before modelling its solution. 
We observed how modelling tools influence students' modelling styles and can be used to provide students with the necessary confidence and creativity. Finally, we formed a set of guidelines based on these observations, which will help in modelling education and help students adhere to the practical skills of modelling. 
The rest of the paper is structured as follows.
%

In Section~\ref{sec:relatedwork}, we discuss the existing literature related to modelling education, students' modelling experience and observational studies. In Section~\ref{sec:method}, we describe the pilot studies, observational study setup, data collection and data analysis strategies. In Section~\ref{sec:results}, we mention the summarised results of the observational study, followed by a detailed discussion of the  results in Section~\ref{sec:discussion}. Finally, we conclude our paper in Section~\ref{sec:conclusion}.

%% file: 02RelatedWork.tex
\section{Related Work}
\label{sec:relatedwork}
Model-Driven Engineering (MDE) or Model-based Engineering (MBE) is the current state-of-the-art in software abstraction, where models are used as primary artifacts in the software engineering process. It has gained popularity in academia and industry for managing the complexity of modern software and is considered a significant advancement in software development \cite{Hutchinson2011Empirical}. 
Although the benefits of MBE are often considered obvious, higher abstraction levels do not guarantee better software and while code generation may boost productivity, the effort to develop models and make manual modifications can negate this benefit \cite{Hutchinson2011MDEpractice}. 
Regardless of the success of MBE, abstract models are crucial to the future of software development. When it comes to software modelling, UML (Unified Modelling Language) and UML-based development methods have become de facto standards in SE. 
Text books on UML and object-oriented design commonly refer to heuristics for creating UML diagrams, e.g., \cite{rumbaugh91,bruegge09,abbott83}.

In the following, we discuss literature related to UML modelling in education, especially modelling done by students, modelling challenges and observational studies as a method of investigation. 
%

\subsection{Modelling in Education and Modelling by Students}
Numerous studies have explored modelling in education, with a predominant focus on teaching modelling methodologies \cite{Westphal2019Teaching,gilson2018teaching,gonnord2018practicing} or presenting experience reports from modelling courses \cite{akayama2013tool,paige2014bad,kolovos2016towards,burgueno2018teaching}. 
In \cite{Westphal2019Teaching}, Westphal presented the design of a software engineering course heavily influenced by and focused on UML modelling. In \cite{gilson2018teaching}, the authors proposed a course where students acted both as language designers and users to evaluate the usability of software language engineering.
Gonnord et al.~\cite{gonnord2018practicing} took a reverse approach, guiding students from low-level C code to designing a modelling language workbench.
%

Akayama et al.~\cite{akayama2013tool} shared their experiences and opinions on tool use in software modelling education.
The paper describes various approaches taken by the individual authors and discusses factors such as modelling tools versus pen and paper, the conflict between design and programming concepts, and methods to measure the quality of models.
Paige et al.~\cite{paige2014bad} discussed what they consider to be bad practices in teaching modelling. They identified issues such as covering too broad a range of modelling-related topics and focusing on syntax instead of semantics.
Similarly, Kolovos and Cabot~\cite{kolovos2016towards} presented a corpus of use cases for courses teaching model-based engineering (MBE). In \cite{burgueno2018teaching}, the authors discuss previous challenges with modelling encountered in various software engineering courses. They then present a case study in which students define a system and its views, simulate them, examine their relationships, and conduct several types of analyses on the complete system specifications. The authors also include a survey to know students' feedback. 

While some surveys assess students' overall experiences with modelling \cite{Ciccozzi2018How,Agner2019Student}, there remains a scarcity of studies providing a detailed account of students' experiences in modelling or reporting results derived from students actively creating models or solving modelling problems. In the domain of business process modelling, studies like \cite{pinggera2013Styles} have documented the modelling experiences of 115 students, highlighting three distinct modelling styles, ``\emph{We could distinguish (1) an “efficient modeling style” characterized by a limited time needed to think about the modeling task, and a fast rate of adding elements to the model; (2) a “layout-driven modeling style” which involves much time in creating a comprehensible layout while being less efficient in creating the model; and (3) an “intermediate modeling style” that is neither particularly efficient nor invests particularly into model layout"}.
%

\subsection{Modelling challenges by students}
Multiple studies reported students' challenges regarding software modelling.
Stikkolorum et al. \cite{Stikkolorum2015LogViz} explore student difficulties and modelling strategies within UML Class diagrams by providing students with a specialized UML editor featuring feedback mechanisms. The findings identify four distinct strategies (depth-less, depth-first, breadth-first, and ad hoc), representing various approaches to constructing class diagrams based on task requirements, encompassing associations, attributes, and problem-solving elements.
Reuter et al.'s research \cite{Reuter2020UMLproblem} delves into students' challenges with UML diagrams across two modelling courses, resulting in the compilation of a comprehensive catalogue of issues associated with UML diagrams. 
Agner et al. \cite{Agner2019Student} broaden the scope with a survey on modelling tool use among 117 students, uncovering issues such as a lack of feedback, diagram-drawing difficulties, and tool complexity. 
In a different study, Hammouda et al. \cite{hammouda2014case} compare the utilization of modelling tools to traditional pen-and-paper methods. Through a survey, they evaluate students' perceptions of the differences, ultimately finding no clear advantage for either approach.
%


\subsection{Observational Studies in Literature}
Observational studies are beneficial in investigating participants' thinking processes and actions. In their paper, Aniche et al.~\cite{aniche2021developers} observe 13 developers thinking aloud different test cases. Their findings show three different strategies used by developers while testing and explain the reason behind these strategies. Dekel et al. \cite{dekel2007notation} perform an observational study of several collaborative design exercises. Their results show the popularity of the ad-hoc nature of collaborative design and the behaviours of different design teams in different contexts and used communication mediums.
In their study, Mangano et al. \cite{mangano2014software} observe eight professional software designers doing 14 hours of design activities on whiteboards to analyse the communication between designers and whiteboard sketching.
Carver et al. report in their study \cite{carver2003observational} an observational approach where an experimental subject executing a procedure was observed by another individual. The study was conducted in pairs, with the participant implementing the technology referred to as the \textbf{executor}, and the individual observing the technology's application termed as the \textbf{observer}. Importantly, the role of the observer did not involve collaborative work with the executor in applying the technology; rather, their responsibility was to ensure that the executor adhered faithfully to the technology's procedural steps. The observer also took notes on the executor's utilization of the technology, highlighting instances where challenges were encountered. After a certain point, the roles of executor and observer were swapped. The authors claimed that this observational study technique provided a level of detail regarding individual process steps and their efficacy that is challenging to obtain through traditional post-experiment questionnaires.
A similar strategy has been used in pair programming for ages. Pair programming \cite{williams2003pair} is becoming a popular practice in software development \cite{williams2003building}.
In their survey, Begel et al. investigate the advantages of pair programming \cite{Begel2008PP}, identifying benefits such as enhanced code understanding, increased creativity and brainstorming, and improved design. 

Motivated by these insights, we embraced a pair programming-style observational study approach to achieve similar benefits in the context of modelling. Acknowledging modelling as an inherently creative task, especially for our predominantly first-year undergraduate participants, collaborative problem-solving emerges as a valuable asset. In contrast to the approach outlined in \cite{carver2003observational}, in our study the \textbf{observer} assumes the role of a \textbf{navigator} who directed the \textbf{executor}. This shift in dynamics differs from the traditional \textbf{executor} role, to a \textbf{driver}, emphasizing the collaborative and interactive nature of our pair programming-inspired methodology.


%% file: 03Method.tex
\section{Method}
\label{sec:method}
Observational studies constitute a cornerstone in scientific research, providing a valuable method for investigating and understanding natural phenomena in their real-world context. Unlike experimental studies, which involve deliberate manipulation of variables, observational studies involve the passive observation of subjects in their natural settings. 
For our study we followed the method proposed by Williams at al.~\cite{williams2003pair,williams2003building} inspired by pair programming, which is a popular practice in software development.
Begel et al. explore the benefits of pair programming in a survey \cite{Begel2008PP}. They identified code understanding, creativity and brainstorming, and better design among the benefits of this method. We decided to follow a pair programming approach (in our case \emph{pair modelling}) for similar advantages. 
We are also following constructive interaction \cite{andriessen2003argumentation,baker1999argumentation,miyake1986constructive} and collaborative learning \cite{veerman1999collaborative} for  data analysis. Furthermore, we are using the Empirical Standards for Software Engineering Research~\cite{EmpiricalStandards20}\footnote{\url{https://www2.sigsoft.org/EmpiricalStandards/tools/}}. In particular we are using author checklist provided in the \emph{General Standard} category, which contains the specific criteria that can be used by authors to conduct and report research.
Understanding and solving a problem with software modelling is a creative task. 
Modellers read the problem from their perspective and conclude the solution based on individual thinking and knowledge of the domain. The modellers' style, apart from modelling tools and notations, influences the solution.
There are no significant studies in Software Engineering (SE) regarding software modelling style. A few studies in business process modelling (e.g., \cite{Pinggera2012Styles}) lead the way towards learning about style. 
%

%
Our study mainly focuses on modelling actions, like \emph{Add, Delete, Edit}, and challenges to understanding students' modelling process. 
We detail our study steps in the following subsections, from the pilot to the final study setup. 

\subsection{Pilot Study 1}
For pilot 1, we utilised the tool Papyrus and a plug-in extension called ModRec \cite{Ragnarsson2021ModRec} to create the models and gather data, respectively. With ModRec, modelling actions can be automatically captured and can be viewed as a separated file. The study was promoted to undergraduate software engineering students at Reykjavik University (RU), resulting in three volunteers. Pilot 1 was conducted in September, 2022. All three participants were briefed on the pilot's objectives and had the option to withdraw at any time. They did not receive any compensation for their participation. It was also clarified that their performance would not have any impact on their grades, neither for good or bad. All three were first-year students familiar with UML modelling.

The students were tasked with drawing a class diagram to depict the domain of a restaurant food court system. They were instructed to use Eclipse Papyrus as their modelling tool and were given a demonstration of the tool's interface and the ModRec plug-in. Several requirements were provided, and they were allotted 30 minutes to complete the diagram.
The details of the problem description used in Pilot 1 can be found in the Appendix~\ref{app:problem_description}.
We observed students facing difficulties with Papyrus, realising they invested most of their time understanding the tool instead of focusing on the task. 
Additionally, we realised that the example domain, a restaurant food court, was not interesting enough to students.

Based on these observations, we made three changes: (i) using a more relevant and exciting problem domain, (ii) adding not just structural, but behavioral aspects to the modelling problem and (iii) not prescribing any modelling tool, letting the participants select one themselves.


\subsection{Pilot Study 2}
After integrating in our study the changes from the pilot study 1, we selected a dating app as the problem domain (more details below) and created two tasks involving drawing a class diagram and sequence diagrams. We introduced structural and behavioural modelling in the pilot 2 to get more details about the students' modelling process. The first task was to draw a class diagram to capture the app's domain. The second task was to draw two sequence diagrams to model two dating app functions.
%

We select the dating app as our domain to get students interested in solving the problem. Our study is based on student volunteers, and getting their attention towards the tasks is necessary. We looked through past courses and realised that it is not a traditional problem domain, and students might be curious to attempt it. 
%
Also, we decided to go with one static and one dynamic UML diagram type. 
Prior experience \cite{chakraborty2023we} suggests that class diagrams are the easiest UML diagram type for students to understand, and sequence diagrams are the hardest. 
Lopes et al. \cite{lopes2019uml} find that students find sequence diagrams not easy to use alone, as they need other diagrams, such as use case and class diagrams. 
In \cite{reuter2020insights}, Reuter et al. survey students' problems with UML diagrams. Regarding sequence diagrams, the authors find students need help with the order of elements, actions in a sequence diagram, needing help to identify interacting classes that are needed as objects for the sequence diagram, etc. 
Inspired by these studies, we decided to use class and sequence diagrams. 
%
Since we let students use their preferred modelling tool, we decided to record their screens to get the modelling actions 
following a method equivalent to pair programming as stated above~\cite{williams2003pair,williams2003building}.

For the first problem, student A plays the driver, and student B be the navigator. That means student A solves the problem by drawing diagrams, while student B navigates through the problem, giving insights and inspecting the drawing. For the second problem, students switched roles.
Also, we thought by pairing up, future participants might feel more motivated to join the study. 
We are aware of the shortcomings of pair programming, such as personality clashes, and bad communication affecting the design. However, we wanted to see how communication between a pair influences the design. 

The study was divided into four sections: Reading the full problem description (10 minutes), one of the participants solving task 1 with a class diagram (20 minutes), taking a small break (10 minutes), the other participant solving task 2 with sequence diagrams (20 minutes).
The details of the problem description used in Pilot 2 can be found in the Appendix~\ref{app:problem_description}.
We then conducted the pilot study 2 in November, 2022 with two PhD students of the computer science department at RU in . Both PhD students had prior experience of UML. The participation was voluntary, they had the option to withdraw at any time and did not receive any compensation for their participation. The two volunteer PhD students selected a tool of their own (draw.io), based on their previous experience of the tool.

After observing the second pilot study, we decided to change the time frame and a few details in the problem description used for the study.  We reduced the time for reading the problem and break to 5 minutes each and added 5 minutes each to the problem-solving. 
We realised that giving participants an open problem description is not a good idea for the problem description, as the two volunteers in the second pilot study took a safe path and created generic diagrams. Hence, for the main study, we added listed requirements to our problem description, which is discussed in the next section.  
For the main study, we recruited volunteers from two universities (11 pairs from University of Malaga, Spain, and 5 pairs from RU, Iceland).

In the next three sections, we discuss the study design and setup (Section~\ref{sec:designandsetup}), data collection (Section~\ref{sec:datacollection}) and data analysis (Section~\ref{sec:dataanalysis}). 

\subsection{Study Design and Study Setup}
\label{sec:designandsetup}

For the final study, we contacted the University of Malaga (UMA) in Spain and RU in Iceland. In UMA, we considered Software modelling and design, a bachelor course in 3rd year with 110 students. The course is about structural modelling with class diagrams and behavioural modelling with sequence diagrams and state machines. It is taught how to model the transition of a system over time. Design patterns and the conversion of class diagrams into object-oriented code are also studied. 
We advertised the study in the course. One of the authors of this paper (a lecturer in the course) presented details of the study and research goals earlier in the course. Later, we asked for voluntary student participants, and 11 pairs signed up for the study. 

In RU, we considered a bachelor course in the second semester of the Software Engineering program, with 51 students. The course centres on analysing, modelling, and designing software systems, employing an object-oriented approach and utilising UML. Throughout the course, students learn class, sequence, state machine, activity, and component diagrams. Additionally, the curriculum includes the practical implementation of the designed systems using the Python programming language.
We made the study the last assignment in the course. Five pairs volunteered for the study. Thus, we asked them to follow the study setup. The rest of the students finish the problem as individual assignments for the course. 

For the final study, we changed the problem description from pilot study 2 and added listed requirements to give students more specific details and restrictions about the problem domain. Then, we kept the tasks similar to pilot 2. 
The details of the final problem description can be found in the Appendix~\ref{app:problem_description}.
Students signed up in pairs, selecting their own partners. 
In UMA, participation was voluntary, and participants could withdraw from the study at any time. Students were orally informed about the purpose of the study: conducting a research experiment. They knew and agreed that their data would be anonymised and the results could be part of research publications. The task was undertaken near the end of the semester, in mid-December 2022.
Students in RU signed a consent form before participating in the study. Participation was voluntary, and participants may withdraw from the study at any time, without penalty. Furthermore, participants could deny the researchers conducting the study the right to use the collected data (from their participation). Since the task was part of the course assignment, participants received grades as per course rules and feedback on their modelling solutions. The study was conducted in April 2023.


Each pair got one laptop and a recorder to record their conversations, and we also asked them to use screen recording so we could record their modelling activities. In UMA, students used Magicdraw as they were familiar with that tool. In RU, students used PlantUML as it was the tool used in that course. Students received the problem description in English. To get their authentic reaction, make them feel comfortable and avoid language barriers, we gave them the option to converse in their mother tongue or in English. All 11 pairs from UMA conversed in Spanish, three out of five pairs in RU spoke Icelandic, and two spoke English. 
%



\subsection{Data Collection}
\label{sec:datacollection}
11 pairs from Malaga participated in the study in Spain. Similarly, five pairs from Reykjavik participated in the study at the end of their course in Iceland. In Malaga, the study was conducted in December 2022, and four months later, in April 2023, the study was conducted in Iceland. 
After a brief analysis of the data collected from all 16 pairs and an extensive analysis with two pairs each from UMA and RU, we concluded the study. We observed a sufficient similarity in the data, indicating a saturation point \cite{baltes2022sampling}. The repetition of data suggested that further collection would be redundant, confirming that we had reached an adequate sample size. Also, we found sufficient diversity, depth and probable issues demonstrating validity of the collected data.
%

Students had 5 minutes to read the problem description, 25 minutes to solve the first problem (class diagram), a break of 5 minutes and then 25 minutes to solve the second problem (sequence diagram). 
We collected three forms of data: (i) \emph{modelling actions through screen recordings}, (ii) \emph{voice recordings of the conversations between a pair}, and (iii)  \emph{PDFs of each pairs' final diagrams}. The three forms of data address our RQs directly. The screen recordings summarise actions taken by students while modelling. All three data sets explain the challenges students face while modelling. 
%

\subsection{Data Analysis}
\label{sec:dataanalysis}
Initially, we classified the recordings (both screen and voice) into three sections: First, reading the problem during the start of the study when a pair reads out the problem (5 minutes); then, drawing the class diagram with a time limit of 25 minutes; finally, drawing two sequence diagrams, also with a time limit of 25 minutes.
We further classified each of these sections into three categories based on our RQs: actions (doing modelling in terms of Adding, Deleting and Editing the diagram), communication (planning modelling) and challenges (about modelling). 
%
While analysing actions, we considered data that describes any modelling action taken by a pair in the modelling tool, e.g., Adding a class, Deleting an attribute, or Adding a connection. We also included timestamps for each of these actions. For communication, we included dialogues between the pair and their timestamps. These dialogues covered agreement or disagreements between a pair, decision-making points, or conversations leading to an action. Finally, challenges include any piece of conversation between a pair or action that describes challenges a pair faces while modelling or about modelling. 
%

We analysed four pairs (2 from UMA and 2 from RU) with the above-mentioned strategy. However, we realised distributing the data like this was tiresome, and threats of researcher bias were increasing. 
Hence, we made a more compact and precise protocol. We focused on the class and sequence diagrams this time, precisely how a pair approached and solved those modelling tasks. 
%

\begin{figure}[t]
    \centering
    \includegraphics[width=0.8\linewidth]{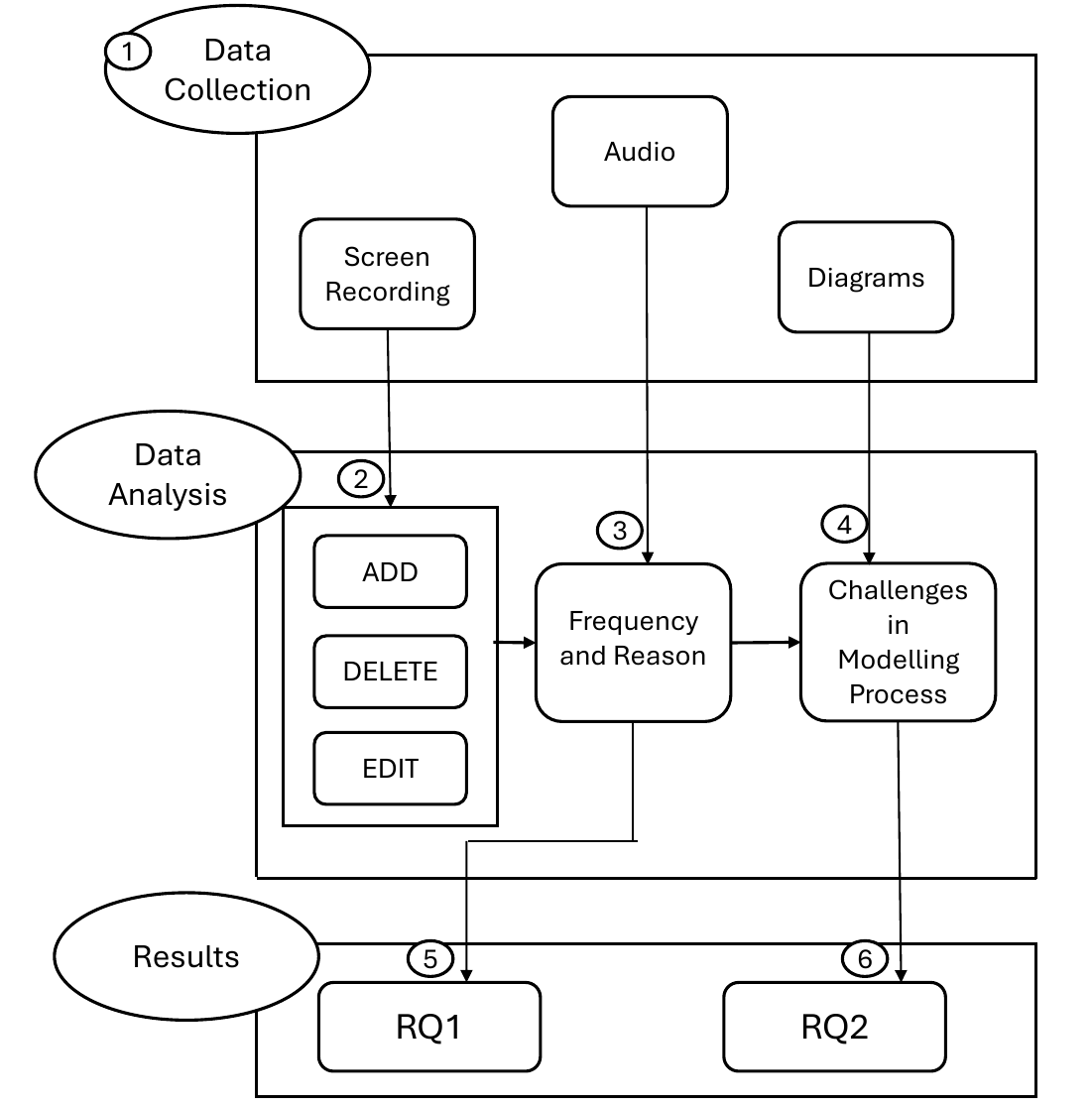}
    \caption{Data Analysis}
    \label{fig:data_analysis}
\end{figure}
Figure~\ref{fig:data_analysis} describes the analysis strategy. We dedicated the first step towards data collection. Next we break down students modelling into three distinct actions, \emph{Add, Delete} and \emph{Edit}, based on the screen recordings. 
An \emph{Add} action occurs when students incorporate new elements into their model, ranging from classes, attributes, and methods to associations between classes, participants, and messages—both direct and return. Conversely, a \emph{Delete} action is logged when students remove any component from their models. Lastly, an \emph{Edit} action encompasses modifications to existing \emph{Add} actions. This involves alterations such as changing the name of a class or message, adjusting multiplicities, and redirecting associations between classes. These defined actions provide a comprehensive framework for analyzing the dynamic evolution of students' modelling throughout the study.

Then, we focused on each action and incorporated the audio feature in the analysis. That is, we tried to understand the frequency of those actions and the reasons behind their occurrence.
%
Finally, we then looked at students' whole modelling process and the final diagrams to understand what type of challenges students experienced while drawing the diagrams. 

Similar to our previous approach, we selected two pairs from each university, performed the analysis, and discussed among the four authors. When we reached an agreement, the first author conducted the rest of the data analysis. 
Based on our complete analysis, we answered the RQs (step 5 and 6). We report them in detail in the next section (Section~\ref{sec:results}).

\subsection{Threats to Validity}
The results in this paper are based on observing 16 pairs of students across two universities in two different countries. The collected data was in three different languages. We identified several threats to validity, which we describe below along with the measures we took to mitigate them, classifying the validity threats in three categories as described in~\cite{wohlin2012experimentation}.%


\subsubsection{Internal validity}
Internal validity focuses on how confident we can be that the elements of the study actually caused the observed outcome. There may be other factors influencing the outcome, which are beyond our control or have not been measured. 

One of the threats in our study is that the students are Icelandic and Spanish, while the problem description was in English. To mitigate this threat, one of the authors, who is Spanish, was present during the study at UMA so that students could clear their doubts. In RU, the original course instructor, who is Icelandic, helped students with questions.

\subsubsection{External Validity}
External validity is concerned with whether we can generalise the results beyond the scope of our study.

Our study participants are all bachelor students, and while doing the study, they were active in a course which involved software modelling with UML. Two threats here are the modelling background of study participants and the use of a UML diagram in the study. For our study setting, it was essential to have a similar modelling knowledge among participants, which was achieved. We used two appropriate modelling diagrams, class and sequence, in the study settings. 

Despite all participants are bachelor students, we believe this threat is minimised because students were active in a course involving software modelling with UML, and they had been doing similar exercises during the course, so all the concepts and knowledge were fresh in their minds. Therefore, we believe the same study involving software engineers might have had similar results. 
This qualitative study, involving 32 individuals, provides valuable insights that can be highly relevant and applicable to similar contexts. While claiming generalisability may not be appropriate, the results offer a strong foundation for understanding the phenomena studied and can potentially inform practices in other comparable settings.

\subsubsection{Conclusion Validity }
Conclusion validity focuses on how confident we can be that the study elements and settings are genuinely related to the observed outcome and whether the study can be repeated. 

We kept a flexible study setting for our student participants as we wanted to capture their interactions with each other and with modelling. The study occurred in their university rooms, where students used their preferred modelling tool and modelled with their language. 
Conclusion validity refers to the extent to which the conclusions drawn from a study are credible and justifiable. It ensures that the relationships identified between variables are genuine and not influenced by external factors or biases. While we cannot entirely eliminate the presence of researcher bias in our results, we implemented multiple measures to mitigate this threat as much as possible. For data analysis, we began with a smaller sample size from our data (two pairs from each university) and had all four authors analyse the data. We then discussed our findings among ourselves, and once we reached an agreement, we analysed the rest of the data accordingly. Consistency was maintained throughout our study design, data collection, and analysis between the two universities. For analysing the screen recordings, we used the terms Add, Delete, and Edit, with all four authors defining and agreeing on these terms. Voice recordings presented a challenge due to the involvement of three languages. However, we transcribed the conversations into English and used live coding to capture the students' perspectives accurately.

%% file: 04Results.tex
\section{Results}
\label{sec:results}
%
We analysed 16 student pairs from two universities as they solved a modelling task by drawing class and sequence diagrams using two modelling tools. We observed several patterns in the actions taken and formulated different challenges faced by the pairs to solve the modelling problem. Below, we answer our two research questions using the collected data. 

\subsection{RQ1: What actions do students take on a modelling tool to solve a task?} 
To answer RQ1, we analysed students' screen recordings and distinguished patterns to draw different diagrams. 
These patterns are on a micro level, detailing how students modelled each element of these UML diagrams. 
Additionally, the patterns observed are specific with respect to class and sequence diagrams and highlight the individuality of students while drawing each type of diagram. 
Combining all the patterns observed from students while drawing both class and sequence diagrams reflects the diverse modelling styles among them. In our previous study \cite{chakraborty2024modelling}, we defined modelling style as \textbf{``consistent individual differences in preferred ways of creating and processing software models"}. The individual preferences displayed while drawing both diagrams revealed hints of distinct modelling styles. In their study, Pinggera et al. \cite{pinggera2013Styles} conducted two large-scale modelling sessions involving 115 students. Their study is similar to ours in terms of recording modelling sessions and considering Add, Edit, and Delete as three distinct modelling actions. After cluster analysis, they reported three distinct modelling styles for business process modelling. We recognise that it is premature to categorise the actions observed in our study as styles based on just two diagrams and 32 students. But unlike Pinggera et al. \cite{pinggera2013Styles}, we recorded students conversation as well, which gave us more depth into their modelling actions. Therefore, we decided to answer RQ1 by presenting these patterns as \emph{``modelling preferences''}, which can later be examined with more diagrams and a wider range of problems.

\subsubsection{Modelling preferences for class diagram}
Students employ diverse strategies when it comes to completing a class with all its attributes and methods, and we observed \emph{three} distinct preferences. \emph{Firstly}, some opt for empty classes, primarily following the ``noun identification'' method. They identify all the necessary classes for the diagram, draw them, and subsequently add attributes and methods. The \emph{second} category involves students starting with the identification of one class at a time, gradually adding basic attributes before progressing to the next class. Lastly, the \emph{third} preference involves a mixed method, where students begin by creating all the classes, then move back and forth between them, adding attributes and methods.
%

By combining the audio with these actions, we observed that students visualised the problem and mapped to its solution in a way that made constructing classes and their relationships appear disorganised. This might be influenced by their experience and familiarity with the diagram style. However, while these factors may play a role, the reason is also related to their personal style. A notable piece of evidence is the difference between the two datasets from the two universities. At UMA, the students were in their third year, while at RU, the students were in their first year. We observed the third approach to drawing classes in four out of the five pairs at RU, which might be due to their inexperience. At UMA, the distribution of the three approaches was almost equal. Hence, we can assume this variation is due to personal style, with individual preferences for moving back and forth between problem and solution, rather than completing a class fully. Instead, they try to complete different pieces of the whole problem and establish the solution in that way.

Students adopt \emph{two} distinct preferences when connecting classes in their class diagrams. The \emph{first} preference involves drawing all classes initially and establishing connections later, while the \emph{second} preference entails sequentially drawing classes and their associations. Variations in the second preference include drawing two classes before connecting them and drawing one class, adding an association, drawing a second class, and then connecting them.

\begin{figure}
    \centering    
    \begin{subfigure}[b]{\textwidth}
        \includegraphics[width=1.0\linewidth]{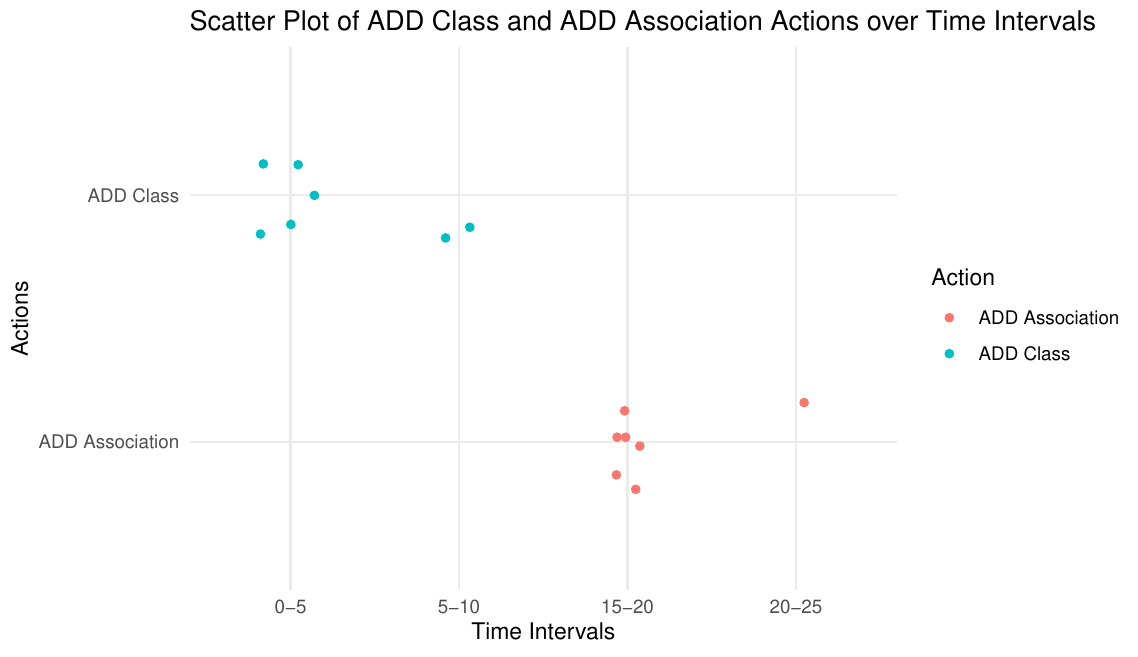}
        \caption{Connecting classes in PlantUML}\label{subfig:connectionclass_RU}
    \end{subfigure}
    
    \begin{subfigure}[b]{\linewidth}
        \includegraphics[width=1.0\linewidth]{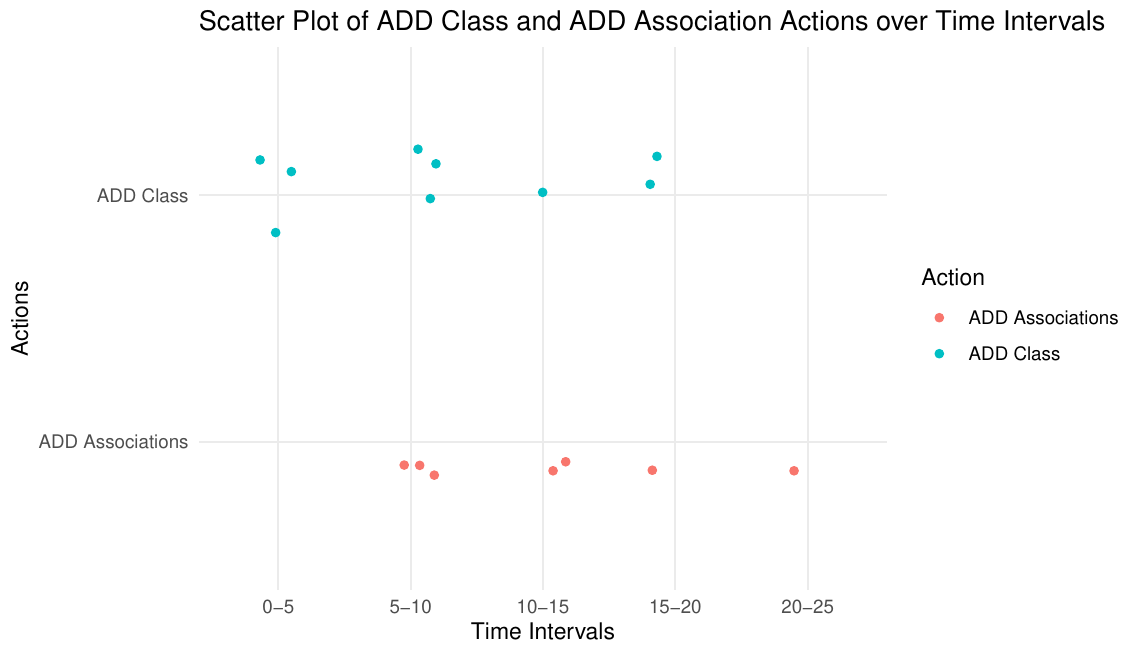}
        \caption{Connecting classes in MagicDraw}\label{subfig:connectionclass_Malaga}
    \end{subfigure}

    \caption{Different ways to connect classes with in PlantUML and MagicDraw}\label{connection_classes}
\end{figure} 
  
Notably, all five pairs using PlantUML, a text-based modelling tool, follow the first preference. In contrast, all 11 pairs using MagicDraw, a more visually-oriented tool, favour the second preference. Hence, it seems the tool significantly influences these choices. PlantUML's code-based nature encourages the construction of larger blocks (classes) first for efficiency, as students can view the generated model after running the code. In contrast, MagicDraw's graphical interface provides visual feedback from the start, with associations drawn by dragging arrows or lines, offering a better visual experience compared to PlantUML. 
Figure~\ref{connection_classes} shows two different behaviours seen in the two modelling tools. In Figure~\ref{subfig:connectionclass_RU}, we can see two actions, Add Class and Add Association, from two pairs of RU students. 
The figure clearly shows how the Add class actions happened initially, followed by the Add association action. In Figure~\ref{subfig:connectionclass_Malaga}, we see a different picture; with the same two actions and two pairs from UMA, we can see the actions are intertwined. 

\subsubsection{Modelling preferences for sequence diagram}
In contrast to class diagrams, students often modified their sequence diagrams. Hence, the emphasis was more on the Delete and Edit actions. 
Students exhibited a higher frequency of Delete actions when working on sequence diagrams, a trend consistently observed across both tools. The increased use of Delete actions suggests a notable pattern in the modification and refinement of sequence diagrams during the modelling process.
After analysing the videos, we concluded the following purposes for using a higher amount of Delete actions in sequence diagrams: 
    \begin{enumerate}
            \item Communication Issues: students were modelling in pairs and following an approach similar to pair programming. We found the \emph{driver} deleting some elements as there was a misunderstanding or miscommunication between the \emph{driver} and the \emph{navigator}. 
            \item Wrong Syntax/Tool issues: the \emph{driver} was realising syntax mistakes and deleting an element, heavily influenced by the tool's interface.  
            When utilising PlantUML, we noted a frequent reliance on the ``copy-paste'' function by students due to its text-based nature. This increased use of the ``copy-paste'' function consequently resulted in a higher frequency of Delete and Edit actions during the modelling process, as pasted content had to be adapted.
            \item Brainstorming: Students deleted elements from the diagram while brainstorming the solution. Students would Add a tentative element, e.g., a class name or an association, to have a basis for discussions. These concepts would then frequently be Edited or Deleted again while the discussion was ongoing.
           
        \end{enumerate}
While reviewing the videos, we encountered situations where Delete and Edit actions were back-to-back, necessitating a clear distinction between these two actions. It is important to note that, in class diagrams, we did not observe this pattern, indicating that students are more confident about modelling the domain with class diagrams.      
%

\begin{figure}
    \centering    
    \begin{subfigure}[b]{\textwidth}
        \includegraphics[width=.9\linewidth]{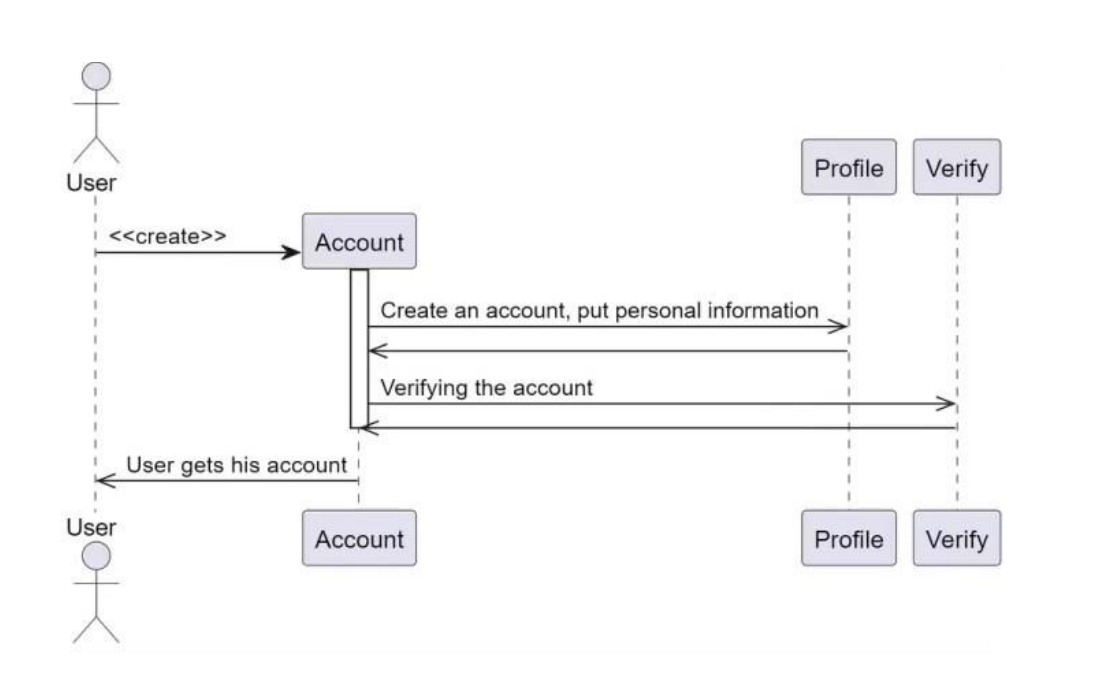}
        \caption{Pair using one direct message line to feed information into the profile}\label{subfig:sd_ru4}
    \end{subfigure}
    
    \begin{subfigure}[b]{\linewidth}
        \includegraphics[width=.7\linewidth]{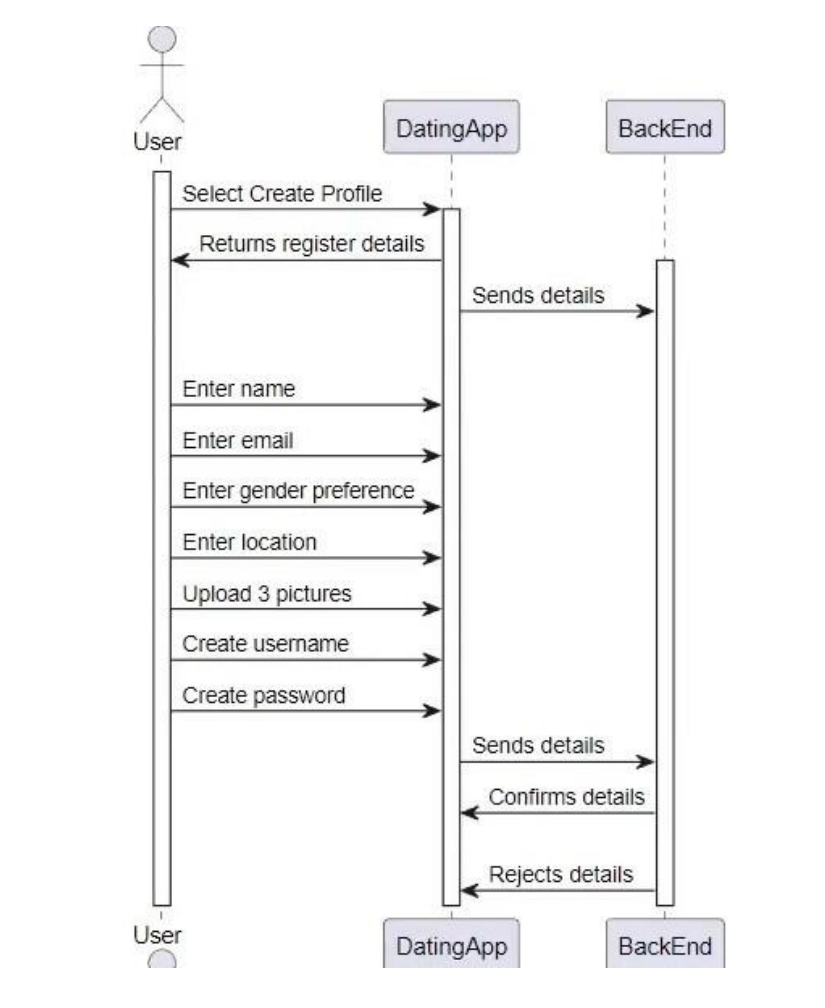}
        \caption{Pair using several direct message lines to feed information into the profile}\label{subfig:sd_ru5}
    \end{subfigure}

    \caption{Sequence Diagrams from RU awith different strategies for profile creation}\label{fig:sd_ru_profile}
\end{figure}

We observed inconsistent preferences or strategies across the sequence diagrams, with students employing varied message types throughout their drawings. We believe this happened because they did not have a clear method, like the ``noun identification'' method, to follow. In Figure~\ref{fig:sd_ru_profile}, we can see two sequence diagrams for creating a profile in the dating app with appropriate information and two different diagrams with different strategies. 
%
In Figure~\ref{subfig:sd_ru4}, students used three objects, i.e., \emph{Account}, \emph{Profile} and \emph{Verify}, to demonstrate profile creation with verification in the dating app. For the same problem, the second pair used two objects \emph{DatingApp} and \emph{BackEnd} in Figure~\ref{subfig:sd_ru5}. Note that both pairs used different strategies to fulfil the requirements, like feeding and verifying information into the profile. Additionally, while the first pair used one direct message line to feed all the personal information into the profile, the second used separate message lines. The problem description listed all the personal information needed to create a profile. Notice the difference between the two pairs and how they read the problem. 

\subsection{RQ2: What challenges do students face while solving modelling tasks?}
To answer RQ2, we integrated the modelling actions with the interactions between students while solving the modelling problem. We observed three main challenges that we detail in the following.

\subsubsection{Inconsistent strategies across different diagrams}
One of the challenges we observed is the use of inconsistent strategies while solving the modelling problem. This issue was less evident in class diagrams. Although we observed different strategies for creating a class diagram, they were not particularly disorganised. However, in the sequence diagrams, students used varying terms to describe messages and objects. 

\begin{figure}
    \centering    
    \begin{subfigure}[b]{\textwidth}
        \includegraphics[width=.9\linewidth]{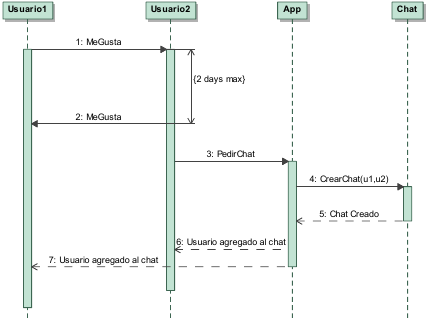}
        \caption{Pair considering the like and the creation of the chat}\label{subfig:sd_mo_uma}
    \end{subfigure}
    
    \begin{subfigure}[b]{\linewidth}
        \includegraphics[width=.9\linewidth]{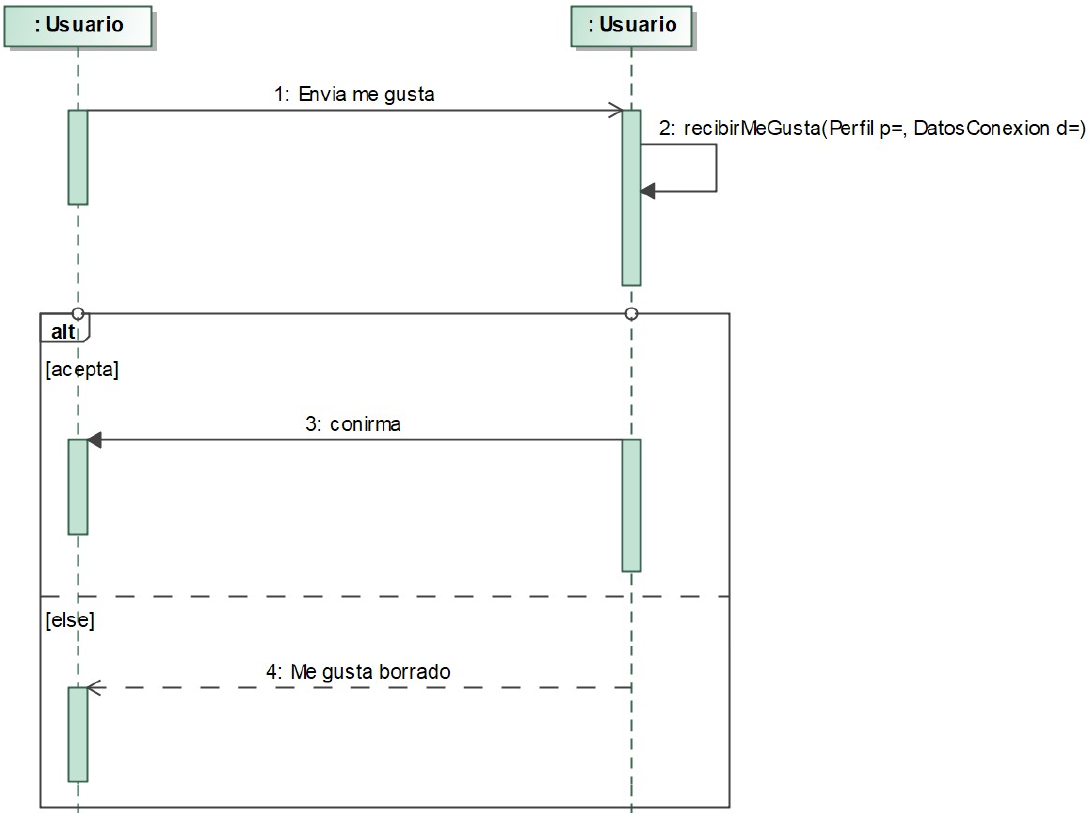}
        \caption{Pair only considering the likes}\label{subfig:sd_sc_uma}
    \end{subfigure}

    \caption{Sequence Diagrams from UMA about giving a ``like'' to someone}\label{fig:sd-uma-like}
\end{figure}

One reason behind this inconsistency is how students interpret the modelling problem. In our previous studies \cite{chakraborty2023we,chakraborty2024evaluation}, we noted that there is no systematic method taught to students for practising the reading of a modelling problem, which is a significant issue.
%
The consequence of this challenge is that students' understanding of the correct diagram can be misleading. As seen in Figures~\ref{subfig:sd_ru4} and \ref{subfig:sd_ru5}, two different diagrams from RU show the same functionality of creating a profile in the app, but each of them employs a different design. A similar situation happens in the diagrams in Figure~\ref{fig:sd-uma-like} from UMA, where we can see that different students go into different levels of detail in the diagrams. We understand that, since modelling is a creative task, having different solutions is predictable. However, guidance is needed to explain to students why they are modelling in different ways.

\subsubsection{Challenges in Interpreting a Modelling Problem}
Our study setting allowed students to read the problem for five minutes before beginning the modelling task. Although students had both tasks (class and sequence diagrams) from the start, the majority focused on reading the problem to draw the class diagram first. Additionally, a common pattern we observed is that students preferred reading the problem description to solve the class diagram and then referred to the class diagram to solve the sequence diagram. Consequently, if they missed something in the class diagram, the subsequent sequence diagram was affected. Students realised the importance of reading the problem thoroughly to solve both tasks only later, while finishing the sequence diagram, and they acknowledged their oversight. 
\interviewquote{N: Let's not do that. We should have fixed that in class diagram, I think it was mentioned in the problem}{RU\_Student\footnote{\textit{D} and \textit{N} refer to \textit{Driver} and \textit{Navigator}, respectively}}

\interviewquote{D: But how can I move that though? Like maybe we have to deactivate. N: Should read first}{RU\_Students}

\interviewquote{D: Can we change the class diagram at this stage (while realising the sequence diagram)? N: I don't think so}{UMA\_Students}

\interviewquote{D: We should have spent more efforts when depicting the class diagram. N: I agree}{UMA\_Students}

\interviewquote{D: I think the thread should have been a class. N: I agree, but that's already done, now we cannot leave it like this in the sequence diagram.}{UMA\_Students}


Following a conversation piece between a pair while solving a sequence diagram task where students had to model the requirements, \emph{a user can ``like" another user if they share the same geographic location, a user has two days to confirm or decline a ``like" from another user"}. This piece of conversation shows struggles and lack of systematic approach to read a modelling problem/model.

\interviewquote{N: It's not like this, because this arrow indicates User1. So maybe we have to put another arrow from User1 to User2 or something. 
D: No, no, it doesn't matter. 
N: Wait, what if both share the same? Oh, no, no. I was reading this wrong. 
D: So okay, so we get a notification we can skip the arrow that goes back to user one, right? 
N: Oh no no, they don't share the same location type}{RU\_Students}


The challenge here highlights students' lack of training in reading problems and diagrams. Having an unsure attempt while drawing such a diagram is understandable. However, a structured way of reading a problem needs to be added here, furthering a method to create this model systematically. We can see students employing both approaches to solve a modelling task, but due to insufficient guidance, they struggle with each.

\subsubsection{Lack of knowledge in the completeness of a diagram}
Students commonly face uncertainty regarding task completion, experiencing doubt after both class and sequence diagram assignments. The challenge lies in the difficulty students encounter in verifying the accuracy of their modelling steps, leading to uncertainty about whether the diagrams effectively address the given tasks. The absence of a clear validation mechanism during the drawing process contributes to this post-drawing doubt among students.


\interviewquote{D: Whatever. We are done, I hope! We just have to fix this later, maybe. N: I think, this is alright(?)}{RU\_Student}

\interviewquote{D: I don't think there is much more we can do.}{UMA\_Student}

\interviewquote{D: There's only 50 seconds left. N: Yeah, I'm not sure about the delete action, let's just leave it like this}{UMA\_Students}


%% file: 05Discussion.tex
\section{Discussion}
\label{sec:discussion}
In this section, we discuss our findings, particularly the issues identified through our results and their implications for modelling education.

\subsection{Different Modelling Preferences}
\begin{figure}
    \centering    
    \begin{subfigure}[b]{\textwidth}
        \includegraphics[width=.9\linewidth]{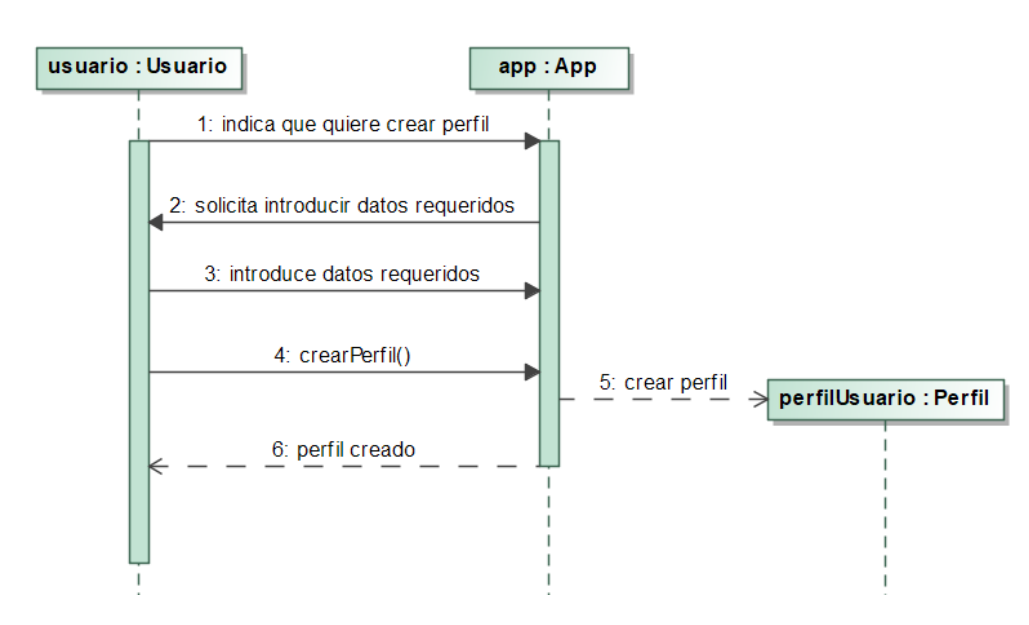}
        \caption{Pair writing mostly sentences in messages}\label{subfig:sd_tl_uma}
    \end{subfigure}
    
    \begin{subfigure}[b]{\linewidth}
        \includegraphics[width=.9\linewidth]{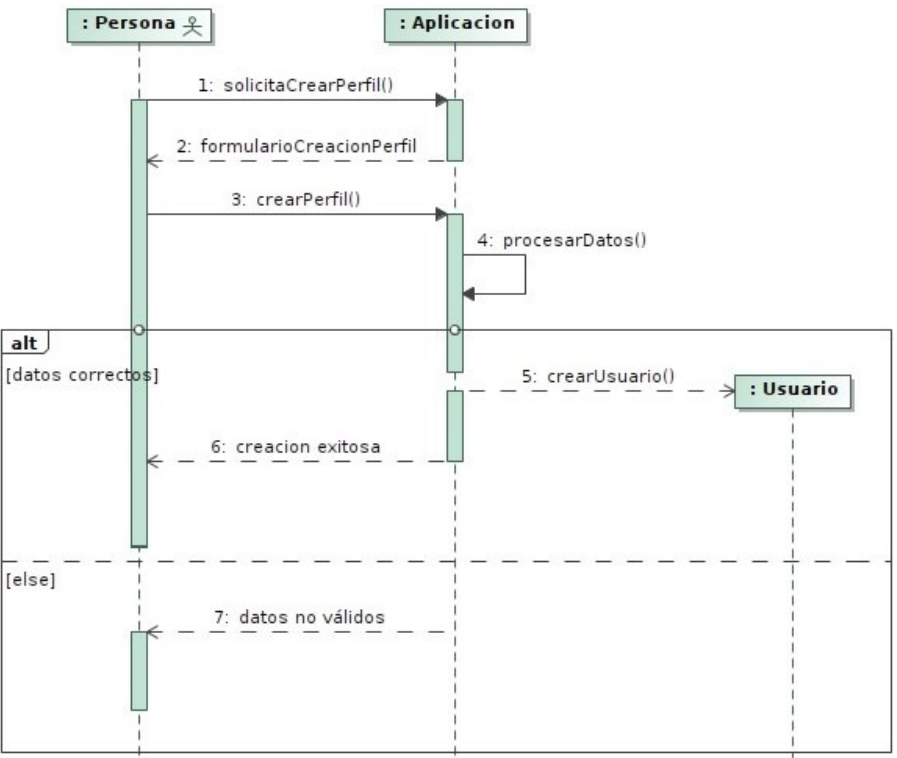}
        \caption{Pair trying to use camel case notation for messages}\label{subfig:sd_aa_uma}
    \end{subfigure}

    \caption{Sequence Diagrams for creating a profile in the app (UMA)}\label{fig:sd-uma}
\end{figure}

Stikkolorum et al. \cite{Stikkolorum2015LogViz} conducted a similar study with 20 bachelor’s students in software engineering during their second semester. Unlike our study, these students were inexperienced with UML modelling and only created class diagrams having learned class diagramming prior to the task. The authors recorded four activities from their class diagram drawing: create, move, set, and remove. They used their own tool, WebUML1, to log these activities. 
%
%
The authors report four strategies for drawing class diagrams in \cite{Stikkolorum2015LogViz}. 
%
We found similar preferences for class diagrams among our students. However, our study, supported by students' conversations and the use of two different tools, provides a more in-depth picture behind these strategies. Three factors influence students' preferences: the modelling tool, the way they read the problem, and their personal style. We observed two main preferences for connecting different classes in the diagram. The first preference involves drawing all classes initially and establishing connections later, while the second preference entails sequentially drawing classes and their associations. 
Notably, all five pairs using PlantUML, a text-based modelling tool, follow the first preference. In contrast, all eleven pairs using MagicDraw, a more visually-oriented tool, favour the second preference. This strongly suggests that the tool has a significant influence on these choices. 
Our study also involves sequence diagrams, which helped us notice different preferences among students based on how they read the problem or the class diagram prior to modelling the sequence diagram.
Additionally, including sequence diagrams shows the difficulties students have in connecting multiple UML diagrams, e.g., by using class diagram elements in their sequence diagrams.
Finally, since we employed a method similar to pair programming, where students alternated their roles, we observed variations in personal style. Some students favoured a more compact layout for the sequence diagram, while many followed the exact wording provided in the problem description, and others chose their own vocabulary. Unlike the class diagram, there was no noun identification method for the sequence diagram. 
In Figure~\ref{fig:sd-uma}, 
we can see two different sequence diagrams for creating a profile in the app, drawn by students from UMA using MagicDraw. 
The pair in Figure~\ref{subfig:sd_tl_uma} mostly writes full sentences for the messages, while the pair in Figure~\ref{subfig:sd_aa_uma} tries to follow the camel case notation and write what could be operation names---although such operations were not added in the class diagram. 



Consequently, individual thinking influenced certain preferences among the students using different tools. In PlantUML, copying and pasting was a common trend, and during brainstorming, we observed that students often paused their actions. In contrast, students using MagicDraw frequently utilised many edit and delete actions, playing around with the interface by dragging parts of the diagram, especially in the sequence diagrams.
An interesting question for future work is whether any of these approaches is more or less suitable for various modelling tasks, such as design exploration.


\subsection{Structural vs Behavioural Modelling}
Students find class diagrams relatively easy to model. Two primary reasons may explain this: the problem description explicitly stated ``model the domain with a class diagram'', which is less complex than modelling two functionalities with a sequence diagram. Another significant reason is that students find more relevance in class diagrams, which they can better visualise in relation to programming. For instance, in RU, the course instructor demonstrated this relevance using programming in lectures. Students often struggle to understand and model the behaviour of the system, finding it more challenging than creating the structural components. The class diagram provides the basic building blocks, which the sequence diagram then has to adhere to. In theory, one could start by designing all the interactions first, making the subsequent class diagram more difficult. However, in practice, students do not start with a ``clean sheet'' when working on sequence diagrams; they frequently refer to the class diagram to construct the sequence diagram.

Various studies \cite{reuter2020insights,lopes2019uml} have addressed challenges with UML diagrams, particularly highlighting the difficulties associated with sequence diagrams. One study mentioned that students find it hard to imagine the sequence of events. In our previous research \cite{chakraborty2023we,chakraborty2024evaluation}, students reported that sequence diagrams were challenging and the sense of a complete diagram was vague. We observed a similar trend in this study; conversations indicated that students felt more confident about finishing the class diagram than the sequence diagram.

\subsection{Influence of Modelling Tools}
Tool-related challenges are not new in software modelling education \cite{Liebel2016Impact,reuter2020insights}. The various designs and interfaces of modelling tools have created complexities in learning modelling. Pourali et al.\cite{Pourali2018DiffTools} show several challenges modellers faced while modelling class and state machine diagrams with eight different modelling tools. 
Our study shows that the tools additionally influence a modeller's modelling style. Our study involved two different modelling tools: MagicDraw and PlantUML. While MagicDraw has a more ``drag and draw'' functionality, PlanUML is code-based. The distinct nature of these two tools influenced our students to follow a certain pattern during modelling. While drawing a class diagram, students using PlantUML created all the classes first and then connected them with the necessary associations. The reason is that, in PlantUML, one needs to run the code in order to get the visual product, and students found it helpful to see all the classes on the screen before completing the connection between them.
In MagicDraw, the tool interface allows one to draw diagrams by selecting elements from the palette and dragging them to the main panel. Since visually, students could see the model in the making, students chose to complete a small part of the class diagram by creating connections between two classes and then moving forward. 

In sequence diagrams, we observed more modification frequency than in class diagrams, i.e., more Delete and Edit actions. The tool's interface influenced here, too. Since PlantUML is code-based, students used a lot of ``copy-paste" and, hence, used Edit actions. Not having consistent terminologies in sequence diagrams and the code-based nature of PlantUML made students more prone to errors. Comparatively, with Magicdraw, there were more Delete actions, as deleting an element or a thread of messages was easy with a click.

\subsection{Implication to Education}
Based on our findings, we emphasise the importance of training students to interpret a modelling problem and a model, to understand the correctness of a model, and to use modelling tools to foster creativity and an individual style. 
Before teaching modelling, it is crucial to incorporate instruction on interpreting both a modelling problem and a model. This practice does not only provide students with a foundational understanding of the modelling concept, but also instils a sense of correctness in their approach. Training students to read a modelling problem helps them gather the necessary information and map a mental model based on the problem and chosen diagram type. 
Studies in the past have demonstrated the difficulties of interpreting models.
Studies like Kuzniarz et al. \cite{kuzniarz2004empirical} and Zayana et al. \cite{zayan2014effects} explore the involvement of stereotypical examples to enhance the comprehension of UML models. Kutar et al. \cite{kutar2002comparison} investigate whether users' understanding of information in sequence diagrams differs from their understanding of collaboration diagrams. 
We have observed students reading the class diagram to model the sequence diagram, highlighting the necessity of an individual's ability to comprehend a model.
When students read a model, they can understand the reverse journey from creation to the problem, identifying mistakes reflected in the model. This knowledge becomes invaluable when students create their own models. Additionally, reading a model imparts valuable insights into the visual aspects of modelling, helping students understand what the end-user can and should infer from the model. Consequently, students develop a nuanced comprehension of the information conveyed through the model, fostering a more comprehensive and insightful modelling process.

\begin{figure}[!ht]
\centering
\includegraphics[width=.3\textwidth]{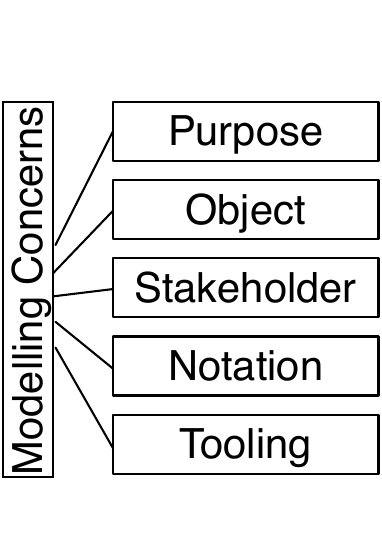}
\caption{Classification of Modelling Concerns. Adapted from \cite{liebel2016model}.}
\label{fig:modelling_concerns}
\end{figure}

Unlike programming, modelling does not provide immediate error feedback. To address this, students should understand when their model is complete and correct. By correct, we mean that the resulting model fulfils all the criteria mentioned in the problem description. A potential approach is to teach students how to interpret a modelling problem into specific modelling concerns \cite{liebel2016model}. Figure~\ref{fig:modelling_concerns} shows five modelling concerns that should be considered while modelling. In the early stages, understanding the objective and purpose is essential. These concerns also assist in interpreting an existing model.
For educators, this implies that these modelling concerns need to be explicit, or exploring the concerns needs to be part of the task, as the solution space otherwise grows.

%

One of the critical revelations from our results is that students' approaches to modelling vary based on the tools they use. Modelling tools are often considered challenging by students \cite{nobrega2007meta,Pourali2018DiffTools}. However, beyond this challenge, modelling tools also shape students' modelling actions. Studies like \cite{bimonte2013design,pati2017proactive,robbins2000cognitive} have investigated and explored different ways a tool can assist modellers; however, these are industry-level studies and do not focus on training students in modelling education. Additionally, the selection of a modelling tool is often left to the students. We believe it is beneficial to let students choose their own modelling tool, but we strongly recommend demonstrating the different interfaces of several modelling tools and allowing students to explore these tools with different assignments. 

%% file: 06Conclusion.tex
\section{Conclusion}
\label{sec:conclusion}
Our study offers insights into students' actions and interactions while solving a modelling problem. 

We conducted a modelling study with 32 students from two universities. The study involved solving a modelling task that included two UML diagrams: class and sequence. Students participated in the study in pairs, and we recorded their screen activities and conversations. After analysis, we observed several modelling preferences and challenges among the participants. These observations revealed modelling preferences that could evolve into distinct modelling styles with more detailed and broader investigation. By examining students' modelling actions alongside their descriptions of the modelling experience, we gained a deeper understanding of modelling as a creative task and identified several issues in modelling education that hinder its later adoption. These issues include inconsistent strategies, a lack of training in interpreting modelling problems, and the absence of methods to verify one's final model. We strongly recommend incorporating guidance into modelling education for interpreting modelling problems and creating models to address these issues effectively. The modelling preferences found through our study are promising and encourage further studies with different diagrams and a diverse range of problems. The results from these studies, in addition to the findings presented here, can lay the groundwork for identifying individual modelling styles, thereby guiding individuals towards more informative modelling practices.


%% file: 07appendix.tex
\section{Problem Descriptions }
\label{app:problem_description}

We present three problem descriptions used throughout our study setup. We changed the details and timings after two pilots. The main study was for 1 hour, divided into Reading the full problem description (5 minutes), one of the participants solving task 1 with a class diagram (25 minutes), taking a small break (5 minutes), the other participant solving task 2 with sequence diagrams (25 minutes).

\subsection{Pilot Study 1}
\begin{itemize}
    \item \textbf{Problem Description}: The task is to model the domain of a restaurant food court system using class diagram.  The system should facilitate the management of multiple restaurants within a food court, customer orders, and the preparation and delivery of food. Your model should accurately represent the relationships between the entities involved and provide a clear structure for the system's operations. Please consider the following requirements for the system:
    The system must support multiple restaurants within the food court,
    each restaurant should have details such as name, cuisine type, menu items, and operating hours, an order should include customer details, a list of ordered items, total price, and order status (e.g., pending, preparing, completed) and finally, the system should support the delivery of orders to customers, including delivery time and delivery personnel details.
    \item \textbf{Modelling Tool}: Please use the Eclipse Papyrus for the task. 
    \item \textbf{Time}: 30 minutes

\end{itemize}

\subsection{Pilot Study 2}
\begin{itemize}
    \item \textbf{Problem Description} The task involves modelling two problems related to an international dating app. Some information of the system: to create a profile, users must provide their personal information, location, and at least three pictures. Every user should have a valid credentials. The app operates by allowing users to see others within a certain geographic area, defined as a flexible distance (x km) from the user's location, which can be either their current GPS location or a manually set coordinate. To initiate a connection, a user can send a "like" to another user within the same geographic area. The recipient will receive a notification with the sender's profile information and has two days to confirm the like. If there is no confirmation within two days, the like is deleted.

    The first task is to model the domain using class diagram. The second task is to model two behaviours of the system: user profile creation and initiating connection between two users. 

    \item \textbf{Modelling Tool}: Please use a modelling tool of your choice.
    \item \textbf{Switching Roles}: After a brief description of pair programming, students were instructed here to switch roles between tasks.

    \item \textbf{Time Limit}: Total 1 hour. Read the problem: 10 minutes, class Diagram: 20 minutes, small break: 10 minutes and sequence Diagram: 20 minutes.
\end{itemize}

\subsection{Main Study}
\begin{itemize}
    \item \textbf{Problem Description}:Your job here is to solve two tasks regarding a dating app. You must maintain the following
constraints while solving the tasks:

1. It is an international dating app, not country specific. \\
2. To open a profile, the user must provide their name, email id, gender preference, location and at
least 3 pictures. Every user should have a username and password. \\
3. How does this app work? Assume User1 and User2 are both users of this app from specific
geographic areas. Two things to notice here: “geographic area” and “user’s location”. A
``geographic area" is a certain distance (x km) from the user's location, which is flexible. A “user’s
location” can be either that user’s current GPS location or some other coordinate set by the user
manually. For example, if User1’s location is Reykjavik, User1 can set his/her/their geographic
location within 80 km of Reykjavik. \\
4. User1 can see all other users of this app with the same geographic location. \\
5. To initiate a connection an interested user of this app send a “like”. User1 can like User2 if both
share the same geographic location. User2 will get a notification, including User1’s profile
information. User2, in this case, has two days to confirm User1's like; a confirmation means User2
also like User1. If there is no confirmation, after two days User1's like will be deleted. \\
6. If both User1 and User2 send like to each other, the app initiates a message thread private to
both users. If none of the users starts talking after 1 week, the thread will be deleted, including
both users' like.

    \item \textbf{Task 1: Class Diagram} Your task is to model the domain of this app, that is, design the domain model using a class diagram.
Please mention the attributes of each class and the multiplicities. Try to think about the operations for each class and mention as many as you can. The purpose here is for a pair to decide what should be in the system and what’s outside of the system. 
    \item \textbf{Task 2: Sequence Diagram} Your task is to model the sequence of the following functionalities: \\
1. User profile creation: Show step-by-step how a user creates a profile in the app. Try to be as detailed as possible. \\
2. Users initiate the connection: Now that users have created their profiles, show step-by-step how a connection is made between two users. (Use the description given above to understand the communication between the system and its users, to draw the sequence diagram)\\

    \item \textbf{Modelling Tool}: Please use a modelling tool of your choice.
    \item \textbf{Switching Roles}: After a brief description of pair programming, students were instructed here to switch roles between tasks.

\item \textbf{Time Limit}: Total 1 hour. Read the problem: 5 minutes, class Diagram: 25 minutes, small break: 5 minutes and sequence Diagram: 20 minutes.

\end{itemize}